# Low-Power Optical Traps using Anisotropic Metasurfaces: Asymmetric Potential Barriers and Broadband Response

N. K. Paul, J. S. Gomez-Diaz*

Department of Electrical and Computer Engineering, University of California Davis, One Shields Avenue, Kemper Hall 2039, Davis, CA 95616, USA

*jsgomez@ucdavis.edu

*We propose the optical trapping of Rayleigh particles using tailored anisotropic and hyperbolic metasurfaces illuminated with a linearly polarized Gaussian beam. This platform permits to engineer optical traps at the beam axis with a response governed by nonconservative and giant recoil forces coming from the directional excitation of ultra-confined surface plasmons during the light scattering process. Compared to optical traps set over bulk metals, the proposed traps are broadband in the sense that can be set with beams oscillating at any frequency within the wide range in which the metasurface supports surface plasmons. Over that range, the metasurface evolves from an anisotropic elliptic to a hyperbolic regime through a topological transition and enables optical traps with distinctive spatially asymmetric potential distribution, local potential barriers arising from the momentum imbalance of the excited plasmons, and an enhanced potential depth that permits the stable trapping of nanoparticles using low-intensity laser beams. To investigate the performance of this platform, we develop a rigorous formalism based on the Lorentz force within the Rayleigh approximation combined with anisotropic Green's functions and calculate the trapping potential of nonconservative forces using the Helmholtz-Hodge decomposition method. Tailored anisotropic and hyperbolic metasurfaces, commonly implemented by nanostructuring thin metallic layers, enables using low-intensity laser sources*

*operating in the visible or the IR to trap and manipulate particles at the nanoscale, and may enable a wide range of applications in bioengineering, physics, and chemistry.*

**1. Introduction**

The optical trapping of small particles in the micrometer range has triggered numerous applications in microbiology [1-3], colloidal dynamics [4], and lab-on-a-chip applications [5], among many others [6-9]. In conventional optical tweezers [10-13], an optical trap is set through a tightly focused laser beam that confines the particle near the higher electric field intensity. There, the gradient of the electric field intensity that surrounds the particle generates the required trapping forces. Unfortunately, it is challenging to extend this approach to trap particles whose size lie down in the nanometer range as (i) gradient forces significantly lessen with the third power of the particle size [14]; and (ii) the thermal fluctuation induced motion of the particles increases [15,16], thus favoring them to escape from the trap. As a result, stable trapping demands high-intensity and tightly focused laser beams that may damage the nanoparticles due to photoheating.

These challenges can be alleviated by exploiting the properties of surface plasmon polaritons (SPPs) [17-20], which are confined electromagnetic waves that propagate at dielectric-metal interfaces [21]. For instance, let us consider an electrically polarizable Rayleigh nanoparticle (with radius $a < \lambda_0/20$, where $\lambda_0$ is the wavelength) located near the surface of a metal and illuminated with light. The particle scatters the incoming light as a superposition of propagative plane waves and evanescent waves. This linear scattering process can be accurately modelled considering that the particle behaves as a polarized electrical point emitter and then using the angular spectrum representation of a source [21]. When the particle is located in the near field of the plasmonic surface, the scattered evanescent waves can couple to the structure and excite guided SPPs [22-27]. Remarkably, this evanescent-wave coupling is governed by spin-orbit interactions [28-30]: only those surface plasmons that possess identical transverse spin to the one of the incoming waves will be excited. In the cases that the particle acquires a linear polarization, the scattered evanescent spectrum lacks any spin and excites SPPs propagating along all directions within the surface. The situation is different when the particle acquires an out-of-plane polarization with respect to the surface, which usually occurs when the incoming beam is circularly polarized [22]. There, the scattered evanescent spectrum acquires a transverse spin and excites only SPPs with similar spin that correspond to plasmons travelling towards a specific direction in the surface. To compensate for the momentum of these directional SPPs, a nonconservative recoil force is exerted on the particle acting in the direction opposite to the plasmons wavevector [22-27]. The direction and strength of this force mostly depends on the handedness of the particle polarization and the momentum of the excited plasmons, respectively [23]. Aiming to boost the strength of recoil forces, anisotropic and hyperbolic (HMTSs) metasurfaces have been proposed to substitute bulk plasmonic metals [31]. HMTSs [32-39] are ultrathin surfaces that exhibit a metallic or dielectric response as a function of the electric field polarization, possess a very large local density of states, and support ultra-confined SPPs over a broadband frequency range. These structures can be constructed by appropriately patterning common plasmonic materials, such as silver [38], gold [39] or graphene [34,35]. It has been shown that the recoil force acting on nanoparticles located over anisotropic metasurfaces can be enhanced several orders of magnitude with respect to the one appearing over bulk, isotropic surfaces [31]. Such giant enhancement is enabled by the large momentum of the directional plasmons excited during the scattering process. Furthermore, the enhancement is broadband [31] in the sense that it appears when the particle is illuminated with light oscillating at any frequency within a very wide range defined by the anisotropic features of the structure.

In this context, recoil forces have recently been exploited to trap nanoparticles near metals using a linearly polarized Gaussian beam [40]. This elegant approach takes advantage of the peculiar distribution of the electric field within the beam: the components parallel to the surface are even-symmetric with respect to the laser beam axis whereas the out-of-plane component is odd-symmetric. The interplay between even/odd symmetries of the in-/out-of- plane electric field components enforces that the nanoparticle acquires an out-of-plane polarization with a handedness that rotates always pointing away from the beam axis and thus excite SPPs towards this direction. This response holds independently of the particle position within the beam. The combination of recoil forces coming the excitation of SPPs in the scattering process together with gradient forces originated from the Gaussian beam generates an optical trap located exactly at the beam axis [40]. Besides, changing the beam focus may reverse the out-of-plane polarization state acquired by the nanoparticle and generate anti-trapping forces that repel it from the beam. However, this platform might not be suitable for many practical applications because it requires specific laser sources operating at wavelengths very close to the metal plasmon resonance. As the laser operation frequency is shifted away from such resonance, the presence of the metal does not play a significantly role on the forces acting on the particle and the platform simplifies to a common optical tweezer governed by gradient forces originated from the Gaussian beam. In addition, the performance of this approach in terms of potential distribution, trap depth, and minimum beam intensity required to achieve stable optical trapping has not yet been investigated. The calculation of these parameters is challenging due to the intrinsic nonconservative nature of the recoil forces that conform the optical trap, which prevents using common theoretical approaches based on the definition of potential energy within conservative force fields [21].

In this contribution, we propose the stable optical trapping of nanoparticles using anisotropic and hyperbolic metasurfaces illuminated with low-intensity Gaussian beams. This platform, illustrated in Fig. 1, permits to engineer optical traps in which giant, nonconservative recoil forces coming from the directional excitation of ultra-confined SPPs determine the overall performance of the trap. The incident Gaussian beam enforces that the nanoparticle acquires an adequate out-of-plane polarization and set the optical trap at its axis. Strikingly, and in stark contrast with the case of bulk metals studied in Ref. [40], the properties of the trap are directly linked with the anisotropic and broadband features of the supported SPPs, and thus they can be modified by tailoring the electromagnetic response of the metasurface. In general, and compared to traps set over common metals, the proposed optical traps exhibit (i) *significantly larger trapping forces*, associated to the high momentum of the supported plasmons; and (ii) *a broadband response*, in the sense that they can be set with beams oscillating at any frequency within the wide range in which anisotropic metasurfaces supports SPPs. In order to investigate this platform, explore its practical viability, and compare its performance with respect to other configurations, we develop below a rigorous theoretical formalism based on (i) the Lorentz force within the dipole approximation merged with anisotropic Green's functions [21] to compute the trapping forces; and (ii) the Helmholtz-Hodge decomposition method [41] to compute the potential energy of nonconservative forces. We validate our results using full-wave numerical simulations in COMSOL Multiphysics [42]. Our approach permits to calculate the spatial potential distribution of the trap, including the trap depth, and allows to elucidate the minimum beam intensity required to achieve stable optical trapping. We have applied our formulation to explore the trapping response of two realistic configurations, namely a silver substrate and an array of densely-packed silver nanostrips [38] that behaves as a hyperbolic metasurface. Numerical results reveal an outstanding trapping performance of nanostructured silver over an ultra-wide frequency range from the visible to the IR. Compared to the case of bulk silver, the nanostructured configuration greatly enhances the trap depth over the entire band and manifold reduces the beam intensity required to achieve stable optical trapping. Even at the silver plasmon resonance, the nanostructured platform exhibit far superior performance than the bulk material due to its large light-matter interactions. We also explore the asymmetrical potential distribution of the trap as the topology of the nanostructure silver layer evolves from elliptical to hyperbolic regimes going through its topological transition, and we reveal the presence of local potential barriers that might appear along precise directions within the surface. Such potential barriers arise due to the momentum imbalance of the SPPs excited over the anisotropic surface and how such excitation changes with respect to the particle position within the beam. These local potential barriers exhibit larger energy than the trap depth and might be useful to predict the direction taken by an energetic particle to escape from the trap. This response is in stark contrast with the symmetrical and smooth potential distributions of traps set over bulk metals. Anisotropic and hyperbolic metasurfaces are promising candidates to trap and manipulate nanoparticles using low-intensity sources operating in the visible and near-IR band, and might lead to important applications in a wide variety of fields ranging from physics and chemistry to bioengineering.

**2. Theoretical formalism: Trapping forces and potential over anisotropic metasurfaces**

This section details first a theoretical framework able to compute the nonconservative optical forces exerted on a dipolar Rayleigh particle located above an anisotropic metasurface that is illuminated by a Gaussian beam. Then, the spatial potential distribution of the trap is computed using the Helmholtz-Hodge decomposition (HHD) method [41]. Our formalism permits to quantitatively determine relevant parameters such as the trap depth and stiffness, trapping forces and potential, and minimum beam intensity required to achieve stable trapping, among others. The approach is general in the sense that no assumptions have been made with respect to the type of metasurface, Rayleigh particle, surrounding media, and operation frequency.

*2.1 Optical trapping forces over anisotropic metasurfaces*

Let us consider an isotropic, non-magnetic, and electrically polarizable spherical Rayleigh particle located at a position $\bar{r}_0 = (x_0, y_0, z_0)$ above an anisotropic metasurface defined by a conductivity tensor $\bar{\bar{\sigma}}^{eff} = \sigma_{xx}^{eff}\hat{x}\hat{x} + \sigma_{yy}^{eff}\hat{y}\hat{y}$, as shown in Fig. 1. The ultrathin metasurface is placed in the plane $z = 0$, lying on the interface between two media with refractive index $n_1$ (top) and $n_2$ (bottom). The particle is illuminated by a normally incident Gaussian beam, i.e., beam axis is aligned with the $\hat{z}$-axis [43], that has a beam width $w_0$ and is focused at a distance $f_0$. The focus position $f_0$ is defined as the vertical distance between the metasurface and the center of the Gaussian beam [43], and it is positive (negative) when the beam is focused above (below) the metasurface. Assuming an $e^{-i\omega t}$ time dependence, the total time-averaged optical forces exerted on the particle are given by [21]

$$\bar{F} = \frac{1}{2}\text{Re}\{\bar{p}^* \cdot \nabla[\bar{E}^{GW}(\bar{r}_0) + \bar{E}^s(\bar{r}_0)]\}. \tag{1}$$

Here, $\bar{p} = \bar{\bar{\alpha}} \cdot \bar{E}^{GW}(\bar{r}_0)$ is the particle's electric dipole moment, $\bar{\bar{\alpha}}$ is the effective dipole polarizability tensor [43], $\bar{E}^{GW}$ is the superposition of the electric field from the incident laser beam and its reflection from the metasurface, and $\bar{E}^s$ denotes the electric field scattered by the particle that are reflected back from the metasurface [21]. Eq. (1) shows that the total force acting on the nanoparticle is composed of two components: (i) the conservative gradient force, $\bar{F}_{grad} = 0.5\, \text{Re}[\bar{p}^* \cdot \nabla \bar{E}^{GW}(\bar{r}_0)]$, that always acts toward the higher electric field intensity of the beam [44,45]; and (ii) the nonconservative recoil force, $\bar{F}_{rec} = 0.5\, \text{Re}[\bar{p}^* \cdot \nabla \bar{E}^S(\bar{r}_0)]$ that appears to compensate the momentum of the directional SPPs excited on the surface [22-25,31]. These two force components have a very different origin: the gradient force depends on the gradient of the electric field intensity surrounding the particle, and thus vary with the type of beam employed. For instance, in the case that of plane waves this term would lead to a radiation pressure pointing toward the direction of the wavefront, whereas in the case of a Gaussian beam this component lead to gradient forces pointing towards the beam center, as in common optical tweezers [12]. On the other hand, the recoil force mostly depends on the properties of the surface plasmons supported by the anisotropic metasurface [31]. Besides, this force also depends on the effective polarization acquired by the particle [31]. For a given distance between the particle and the metasurface, the recoil force is maximized (strictly zero) when the particle acquires an out-of-plane circular (linear) polarization.

The electric field of the p-polarized Gaussian beam employed in the proposed platform possesses x and y components (in-plane) that are even-symmetric with respect to the beam axis, whereas the z component (out-of-plane) is odd-symmetric [40]. Such electric field polarizes the nanoparticle with an out-of-plane polarization state that is independent of the particle position within the beam [43]. When the particle is located close to the metasurface, the total electric field acting on it also depend on the fields reflected on the surface. The resulting non-paraxial electric field components yield [21]

$$E_x^{GW}(\bar{r}) = \frac{w_0^2}{4\pi} \iint_{-k_1}^{k_1} \left\{ \frac{k_x k_{z1}}{k_t k_1} e^{-ik_{z1}z} - \left( R_{sp} \frac{k_y}{k_t} + R_{pp} \frac{k_x k_{z1}}{k_t k_1} \right) e^{ik_{z1}z} \right\} e^{-\frac{k_t^2 w_0^2}{4}} e^{ik_{z1}f_0} e^{i(k_x x + k_y y)} dk_x dk_y, \quad (2a)$$

$$E_y^{GW}(\bar{r}) = \frac{w_0^2}{4\pi} \iint_{-k_1}^{k_1} \left\{ \frac{k_y k_{z1}}{k_t k_1} e^{-ik_{z1}z} - \left(R_{sp}\frac{k_x}{k_t} - R_{pp}\frac{k_y k_{z1}}{k_t k_1}\right) e^{ik_{z1}z} \right\} e^{-\frac{k_t^2 w_0^2}{4}} e^{ik_{z1}f_0} e^{i(k_x x + k_y y)} dk_x dk_y, \quad (2b)$$

$$E_z^{GW}(\bar{r}) = \frac{w_0^2}{4\pi} \iint_{-k_1}^{k_1} \frac{k_t}{k_1} \left\{ e^{-ik_{z1}z} + R_{pp} e^{ik_{z1}z} \right\} e^{-\frac{k_t^2 w_0^2}{4}} e^{ik_{z1}f_0} e^{i(k_x x + k_y y)} dk_x dk_y. \quad (2c)$$

Here, $k_1$ is the wavenumber in the medium above the surface with a transverse component $\bar{k}_t = \hat{x}k_x + \hat{y}k_y$ and a vertical component $k_{z1} = \sqrt{k_1^2 - k_t^2}$; and $R_{pp}$ and $R_{sp}$ are the Fresnel reflection coefficients that characterize the reflection of 'p' (transverse magnetic, TM) and 's' (transverse electric, TE) waves from the anisotropic surface when it is illuminated with 'p' waves [43]. In addition, a phase shift $e^{ik_{z1}f_0}$ is introduced as a measure of tuning the laser focus position $f_0$ along the $\hat{z}$-axis [21,40]. Note that the integration limits in Eq. (2) are set to $\pm k_1$ because propagative modes dominate the response of the beam and the evanescent spectrum is negligible [40]. In most scenarios, the total fields described in Eq. (2) keep a similar symmetry as the incident Gaussian beam in free space and polarize the particle with the desired handedness. It should be noted that the symmetry of these fields may change when the Gaussian beam is focused well below the metasurface. In that case, described below, the particle may acquire an out-of-plane polarization that rotates pointing toward the beam axis and the recoil forces become "anti-trapping" forces [40].

From Eq. (1), the lateral components of the gradient and recoil forces can be simplified as [43]

$$\bar{F}_{lateral,grad} = \frac{1}{2} \text{Re} \sum_{n=x,y,z} \left[ p_n^* \frac{d}{dx} E_n^{GW}(\bar{r}_0) \hat{x} + p_n^* \frac{d}{dy} E_n^{GW}(\bar{r}_0) \hat{y} \right], \quad (3a)$$

$$\bar{F}_{lateral,rec} = -\omega^2 \mu_0 \left[ \text{Im}[p_x^* p_z] \text{Im} \left[\frac{d}{dx} G_{xz}^s(\bar{r}_0)\right] \hat{x} + \text{Im}[p_y^* p_z] \text{Im} \left[\frac{d}{dy} G_{yz}^s(\bar{r}_0)\right] \hat{y} \right]. \quad (3b)$$

Eq. (3a) shows that the gradient force always acts toward the maximum electric field intensity (i.e., toward the beam axis) of the standing wave formed by the incident and reflected fields above the metasurface. In addition to the type of beam, this force also depends on the particle's polarizability [21]. Eq. (3b) shows that the direction of the recoil force is determined by the interplay between the particle's in-plane ($p_x$ and $p_y$) and out-of-plane ($p_z$) dipole moment components. Using a properly focussed Gaussian beam, the particle acquires an out-of-plane polarization that rotates against the beam axis associated with recoil forces directed towards the beam. In case of isotropic metasurfaces, these forces point exactly towards the beam axis independently of the particle position within the beam [40]. In case of anisotropic metasurfaces, the direction of the recoil force may not point towards the beam axis due to the broken out-of-plane symmetry of the system [i.e., $G_{xz}^s(\bar{r}_0) \neq G_{yz}^s(\bar{r}_0)$ in Eq. (3b)]. As discussed below, the recoil force will then push the particle towards the beam axis following a parabolic trajectory. In addition, Eq. (3b) unveils that the strength of the recoil force depends on the imaginary part of the spatial derivative of scattered Green's function out-of-plane tensor component, which measures the momentum of the excited directional plasmons [23,31].

An important parameter that defines the performance of an optical trap is the trap stiffness, which measures the restoring force that acts on the nanoparticle to bring it back to a stable position within the trap –similar to the spring constant in a common mechanical system. This parameter is more significant in Brownian systems, where particles suspended in liquids may acquire random motion due to the continuous collision with the moving fluid molecules. The stiffness of a trap set over a surface can be computed as [46]

$$\kappa(\phi) = -\frac{F_\rho(\rho,\phi)}{\rho}\bigg|_{\rho\to 0} , \qquad (4)$$

where $F_\rho(\rho,\phi)$ denotes the radial component of the lateral forces evaluated at a position $(\rho,\phi)$ defined in polar coordinates with respect to the beam axis. We remark that the polar component of the lateral forces, $F_\phi(\rho,\phi)$, do not contribute to the trap stiffness as it is directed around (instead of towards) the trap. In most cases considered in the literature [12,40], for instance within the forces generated by Gaussian beam in free-space or over common plasmonic materials, the trap stiffness is isotropic in the sense that it has polar symmetry and therefore provides an identical response in all directions: $\kappa(\phi) = \kappa$. This is very different in the case of traps set over anisotropic metasurfaces: the restoring force that a nanoparticle experiences towards the trap depends on the direction from which the particle is trying to escape. Traps with anisotropic stiffness are useful to predict the probable direction followed by the particle when it acquires enough energy to escape from the trap.

*2.2 Trapping potential over anisotropic metasurfaces*

The trap potential is arguably the most important parameter that defines the performance of an optical trap [17,18,47]. Here, we will focus on the trap potential energy and trap depth, which is a quantitative measure of how long the particle remains confined within the trap. In case of conservative forces, such as the gradient force originated within a Gaussian beam [13], the trapping potential U of a vector force $\bar{F}_c$ can be obtained as $U_c(\bar{r}) = -\int_{-\infty}^{\bar{r}} \bar{F}_c(\bar{r}') \cdot d\bar{r}'$ [21]. This potential represents the energy required to move a particle from a reference location with zero energy (taken here in the infinite) to the position defined by the vector $\bar{r}$. Since the vector force is conservative, the path chosen in the integral is not relevant: any trajectory from infinite to $\bar{r}$ provides identical potential energy. This situation is different in the case of nonconservative vector forces because choosing alternative paths might lead to different potential energies, which is the case of recoil forces appearing over plasmonic surfaces [48]. In the case of the platform proposed in Fig. 1, the intrinsic anisotropy of the metasurface makes evident that chosen different paths to move the particle from infinite to the position $\bar{r}$ would require different energies. In all cases, the solenoidal component of nonconservative forces prevents using the classical definition of potential energy. [18]. Here, we apply a different technique to compute the trapping potential based on the Helmholtz-Hodge decomposition (HHD) method [41]. First, we express the force field as [48-50]

$$\bar{F}(\bar{r}) = -\nabla U + \nabla \times \bar{A}, \tag{5}$$

where $\nabla$ is the vector gradient, U is the potential energy, $\bar{A}$ is the vector potential, and $\nabla U$ and $\nabla \times \bar{A}$ denote the conservative and nonconservative force components, respectively. Taking the divergence on Eq. (5) and applying the identity $\nabla \cdot (\nabla \times \bar{A}) = 0$ permit us to find the potential energy through the differential equation [48]

$$-\nabla^2 U = \nabla \cdot \bar{F} \text{ on } \Omega, \tag{6}$$

that is subjected to the Neumann boundary conditions [51]

$$\nabla U \cdot \hat{\rho} = \bar{F} \cdot \hat{\rho} \text{ on } d\Omega, \tag{7}$$

where $\hat{\rho}$ is a unit vector pointing outwards with respect to the boundary of the domain $\Omega$. This numerical approach is valid when the force field is defined over a bounded region $\Omega$ with a smooth boundary condition $d\Omega$. We stress that the platform considered here fulfils these conditions: the domain is defined by the Gaussian beam impinging over the metasurface and the boundary conditions are related to the negligible force acting on the particle when it is located very far away from the beam axis.

We will explore the potential distribution of optical traps set using Gaussian beams over bulk materials and reveal that they are defined by a spatially symmetric function centred at the beam axis. In stark contrast, the trapping potential over anisotropic metasurfaces illuminated with a Gaussian beam lacks polar symmetry with respect to the trap centre. In both cases, the trap depth $\delta_d$ is unique and defined as the potential difference between the beam axis and a position located in the infinite with zero energy. Strikingly, and as further detailed below, the intrinsic anisotropy of the metasurface gives rise to *local potential barriers* with larger energy than the trap depth. As a result, the particle might acquire enough energy to escape from the trap but not to overcome such potential barriers and thus will follow a special route within the plane to avoid them. Finally, it should be noted that stable optical trapping appears when the trap depth is larger than $10k_B T$, where $k_B$ is the Boltzmann constant and T is temperature. If this condition is not fulfilled, mechanisms such as thermal fluctuation [52,53] and Brownian motion [12,17,50] may provide enough energy to the particle to quickly escape from the trap. Thus, the minimum laser beam intensity required to achieve stable trapping is the one required to generate an optical trap with a potential depth $\geq 10k_B T$ [21].

## 3. Performance of optical traps engineered over anisotropic and hyperbolic metasurfaces

In this section, we explore the performance of optical traps engineered over anisotropic and hyperbolic metasurfaces illuminated by a p-polarized Gaussian beam. To this purpose, we first analyze the recoil and gradient forces acting on a nanoparticle versus its position with respect to the beam axis, unveiling the mechanisms that conform the optical trap. Then, we investigate key parameters of the trap including spatial potential distribution, trap depth and stiffness, local potential barriers, and the laser beam intensity needed to achieve stable trapping versus the wavelength of the incoming beam. As the wavelength increases, the metasurface topology evolves from an anisotropic elliptical to a hyperbolic regime going through a topological transition, which permits to study how the different light-matter interactions enabled by these regimes conform the properties of the optical trap. During our study, we compare the performance of the proposed traps to the one found using Gaussian beams in free-space [11-12] and over common plasmonic surfaces [40], aiming to highlight the pros and cons of this platform with respect to other configurations and assess its practical viability.

In the following, we consider a spherical gold nanoparticle of radius $a = 15$nm located at $\bar{r}_0 = (x_0, y_0, a)$. The metasurface is constructed using nanostructured and periodic silver rods [38] with width W= 60nm, height H = 10nm and periodicity L = 180nm (see Fig. 1) patterned over a porous polymer with refractive index $n_2 = 1.05$ [25]. The subwavelength thickness and periodicity of the layer allow us to characterize it using an effective in-plane conductivity tensor [54-56] with negligible out-of-plane polarizability [57]. Even though the use of different substrates might change the particle polarizability and the density of states provided by the structure, the overall response will not be significantly affected [43]. We have carefully verified the accuracy of our model using full-wave numerical simulations as well as the dispersive hyperbolic response of the surface [43]. For comparison purposes, we employ bulk silver with identical properties as the one employed on the nanostructured metasurface [43].

*3.1 Optical forces arising in anisotropic traps*

Fig. 2 illustrates the response of the proposed optical trap detailing the different forces that act on the nanoparticle when it is illuminated with a Gaussian beam at 540nm. At this wavelength, the nanostructured silver layer behaves as a hyperbolic metasurface [43]. For the sake of simplicity, we begin considering that the nanoparticle is located along the metallic rods (i.e., the $\hat{x}$-axis). In this situation, the incident beam enforces that the particle acquires an out-of-plane xz polarization given by the dipole moment $\bar{p}(x_0)=[p_{xr}(|x_0|) + ip_{xi}(|x_0|)]\hat{x} + [\mp p_{zr}(|x_0|) \mp ip_{zi}(|x_0|)]\hat{z}$, where the subscripts 'r' and 'i' denote the real and imaginary components of a complex number, and the upper (lower) sign appears when the particle is located in the negative (positive) portion of the $\hat{x}$-axis [43]. We stress the symmetry of the electric dipole magnitude with respect to the beam axis, i.e., $|\bar{p}(x_0)| = |\bar{p}(-x_0)|$. This dipole can be expressed as a linear combination of two fundamental emitters that have opposite out-of-plane polarization handedness with respect to the surface. The dipole moment of these emitters are $\bar{p}_1(x_0)=p_{xr}(|x_0|)\hat{x} \mp ip_{zi}(|x_0|)\hat{z}$ and $\bar{p}_2(x_0)=ip_{xi}(|x_0|)\hat{x} \mp p_{zr}(|x_0|)\hat{z}$. In most scenarios, the field distribution of the p-polarized Gaussian beam and its reflection on the metasurface ensure that the dipole $\bar{p}_1(x_0)$, associated with a polarization handedness rotating against the beam axis, is strongly excited and dominates the scattering process [43]. Fig. 2(a) shows the power of the SPPs launched on the metasurface for several particle positions. When the particle is located away from the beam axis (i.e., $x_0 \neq 0$), it mostly scatters evanescent waves with a transverse spin that excites directional plasmons with wavevectors pointing away from the beam axis, associated with a "trapping" recoil force acting towards the beam axis. When the particle is located exactly on the axis of the Gaussian beam, it acquires a linear polarization $\bar{p}(x_0 = 0) = p_x(x_0 = 0)\hat{x}$ and scatters waves without any specific spin that excites SPPs propagating symmetrically through the surface. As a result, the recoil force vanishes, and an optical trap is set at $x_0 = 0$. It is important to note the role of the dipole $\bar{p}_2(x_0)$: it excites directional plasmons propagating towards the beam axis that result into "anti-trapping" recoil forces [40]. In the case shown in Fig. 2, the magnitude of this emitter is very small [43] and thus it barely contributes to the excitation of SPPs. In a more general case, it is possible to engineer trapping or anti-trapping recoil forces by controlling the strength of the orthogonal dipoles that characterize the electromagnetic response of the particle. This can be done by manipulating the properties (focusing, polarization, etc.) of the incident Gaussian beam.

The total optical forces exerted on the nanoparticle are determined by the superposition of gradient and recoil forces. Fig. 2(b) shows the $x$-component of the total (blue solid line) and recoil (red solid line) forces versus its position along the $\hat{x}$-axis. Results confirm that giant recoil forces, enabled by the large momentum of the supported SPPs, dominate the response of hyperbolic traps whereas gradient forces play a moderate role. Numerical full wave simulations performed in COMSOL Multiphysics (markers) are included for validation purposes. For the sake of comparison, the panel also shows the forces arising when the nanostructured layer is replaced by pristine bulk silver (dashed lines). It should be noted that, at this wavelength, silver barely interacts with the incoming light and only supports weakly-confined plasmons [43]. As a result, the gradient force coming from the Gaussian beam clearly dominates over the recoil one on the optical trap whereas the influence of silver is negligible: it basically behaves as a reflector that helps to better polarize the particle. Overall, the hyperbolic response of nanostructured silver enhances the force strength over six times with respect to the unpatterned case. This example highlights how anisotropic metasurfaces can enable plasmon-assisted optical traps at desired wavelengths determined by the surface properties. Note that silver can provide significant recoil forces when it is illuminated by a beam oscillating close to the material plasmon frequency, a case studied in detail below. Fig. 2(c) compares the vertical forces acting on the particle when it is located over these two configurations. Above nanostructured silver, the total vertical force is dominated by the recoil force, which is always attractive, pushes the particle towards the surface, and exhibits a maximum strength near the trapping position. In the case of bulk silver, the vertical force is again dominated by the gradient force, which is always repulsive and repels the particle from the surface towards the nearest intensity hotspot [43].

Even though our study above has been focused on nanoparticles located along the metallic rods of the nanostructure ($\hat{x}$-axis in the coordinate system of Fig. 1), the underlying mechanisms led by the particle polarization spin hold independently of the particle position within the surface [43]. Fig. 3 explores this scenario and shows the components of the lateral forces acting on the particle as well as a quiver plot indicating the force direction. Results clearly confirm that an optical trap is created exactly at the beam axis. Furthermore, this analysis also reveals the intrinsic anisotropy of the metasurface: the strength of the recoil force exerted on the nanoparticle lacks radial symmetry. This asymmetry appears because SPPs travelling towards different directions within the surface possess different momentum and spin, as it is evident from Eq. 3(b), and the resulting force might not be directly directed towards the beam centre. Instead, the particle would follow a parabolic trajectory towards the trap, as shown in Fig. 3(c). Note that the recoil force is significantly larger than gradient force for all particle positions and thus determines the trap performance.

*3.2 Performance of anisotropic optical traps versus wavelength*

Fig. 4(a) shows the potential depth of the traps engineered over nanostructured silver versus the wavelength of the incident Gaussian beam. Results have been normalized with respect to the beam intensity. This figure highlights *the extreme bandwidth in which optical traps with very large potentials can be set*, covering the band from around 300nm to over 2$\mu$m, and how the trap depth correlates to the metasurface topology. Theoretically, the structure exhibits hyperbolic responses in the near-infrared and beyond. However, due to the difficulty to appropriately focus the beam at these frequencies due to the diffraction limit as well as the smaller amount of power scattered by the particle there, we will restrict our analysis later to the visible portion of the spectrum. It should be noted that different type of anisotropic and hyperbolic metasurfaces can be designed to operate in the IR region [58-61]. Fig. 4(a) also shows the trap depth obtained with a similar Gaussian beam focused in free-space (black solid line) and over bulk silver (blue solid line). In the former case, the trap depth increases as the laser wavelength decreases, a response associated the higher amount of power scattered by an electrically larger particle. In the latter, bulk silver enables optical traps with maximum potential depth at around 340nm, a wavelength that corresponds to the material plasmon resonance frequency where silver supports moderately confined surface plasmons. Fig. 4(b) shows the minimum laser intensity required to achieve stable trapping (i.e., a trap depth $\sim 10 k_B T$) in these configurations. The study reveals that the nanostructured metasurface permits reducing almost one order of magnitude the required beam intensity with respect to the other platforms. This has significant practical implications as it allows using low-intensity laser sources operating in the visible and potentially the IR to trap and manipulate nanoparticles using anisotropic metasurfaces while avoiding delicate alignments between the surface response and the laser wavelength.

In order to further explore the response of the proposed traps versus wavelength, we study below how light-matter interactions enabled by the anisotropic metasurface, expressed here in terms of topology and wavelength-dependent isofrequency contours (Fig. 5), conform the potential distribution of the trap (Fig. 6 and 7). The potential energy is computed varying the particle position $(x_0, y_0)$ over the surface with respect to the beam axis. We begin our study considering the response of the metasurfaces at $\lambda_0 = 300$nm. At this frequency, silver does not exhibit a plasmonic response (i.e., $\text{Re}[\varepsilon_{r,s}] \geq 0$ [43]) but a dielectric one that does not support confined surface plasmons. As a result, both surfaces behave as dielectric reflectors and the trap depth shown in Fig. 4 is originated by the interference field pattern between the incident and reflected Gaussian beams. At the silver plasmon resonance, found at $\lambda_0 \approx 340$nm [43], the bulk material supports TM isotropic surface plasmons (Fig. 5b) that lead to a radially symmetric potential distribution around the beam axis (Fig. 6b). At this frequency, the nanostructured silver layer behaves as an elliptical anisotropic surface (Fig. 5a) and supports confined surface plasmons. Interestingly, the intrinsic metasurface anisotropy translates into a nonsymmetric potential distribution that is illustrated in a 3D fashion in Fig. 6a. Fig. 7 further studies the 1D potential distribution of this configuration when the particle is swept over the main axes of the beam. Along the silver nanorods (i.e., $\hat{x}$ axis with $y_0 = 0$), the potential is spatially smooth, and the trap depth corresponds to the potential difference between the position with minimum energy at the beam axis and infinite ($\nabla U_x = \delta_d$). Across the strips (i.e., $\hat{y}$ axis with $x_0 = 0$), the potential presents local maxima with energy larger than zero that leads to local barriers with potentials greater than the trap depth ($\nabla U_y > \delta_d$). In this configuration, the potential of the local barrier is comparable to the trap depth because the metasurface is weakly anisotropic [43]. Local potential barriers appear due to the difference on the nonsymmetric and nonconservative recoil forces that act on the particle versus its position within the beam and thus cannot exist in symmetric platforms as the one enabled by bulk silver (Fig. 6b). Remarkably, barriers with potential energies even larger than the trap depth can be obtained by leveraging extreme anisotropy responses, associated with SPPs with drastically dissimilar wavenumbers as they travel towards different directions within the plane. This case can be found at the metasurface topological transition, which appears at $\lambda_0 = 390$nm [43] and exhibits a canalization-like response along the $\hat{y}$ direction [62]. There, plasmons propagating towards the $\hat{x}$ axis possess significantly larger momentum than those traveling toward the canalized direction, enabling local potential barriers along the strips (see Fig. 7b) with an energy $\nabla U_x > \nabla U_y = \delta_d$. In such configuration, a trapped particle that gains kinetic energy will probably escape in the direction perpendicular to the strips, which in addition to lower potential also exhibits a reduced trap stiffness. It should be noted that the trap depth at this wavelength slightly decreases (Fig. 4a) due to the overall moderate local density of states exhibited by the metasurface (Fig. 5a). However, the trap depth is still larger than the one found over bulk silver (Fig. 6b), a material that quickly reduces its light-matter interactions and the momentum of the supported SPPs when operated off-resonance. As the operation wavelength further increases, the nanostructured silver layer behaves as a hyperbolic metasurface and supports highly-confined surface plasmons. Isofrequency contours of these SPPs and associated trapping potentials at $\lambda_0 = 540$nm and $\lambda_0 = 785$nm are shown in Figs. 5-7. Hyperbolic surfaces lead to asymmetric potential distribution and very significant trap depths, greatly extending the functionality of the proposed anisotropic platform from the visible towards the infrared. Local potential barriers also arise in the hyperbolic case due to the different features of plasmons propagating towards *x* (see Fig. 2a) and *y* semi-planes. SPPs properties evolve as the wavelength increases and the metasurface hyperbolic branches slowly close and tend to behave as in a canalization regime along the $\hat{x}$ direction, which in turn leads to local potential barriers across the strips (i.e., $\hat{y}$-axis). For comparison, bulk silver behaves as a lossy dielectric reflector when operated out of resonance and evolves with the higher wavelengths towards a lossy metallic reflector. At these frequencies, bulk silver does not effectively contribute to conform an optical trap rather than enhancing/decreasing the gradient forces acting on the particle by creating standing wave field patterns between the incident and reflected Gaussian beams.

To complete our study, Fig. 8 shows the stiffness of the optical traps engineered over the considered platforms versus the polar angle ϕ within the surface –defined with respect to the positive $\hat{x}$-axis, i.e., along the strips– and the beam wavelength. In the case of the nanostructured silver layer, the trap stiffness dramatically increases when the metasurface topology changes from elliptical TE to anisotropic elliptical TM, at around 340nm. As happen with the potential, the stiffness exhibits a nonsymmetric distribution and, starting from the topological transition at 390nm to around 750nm, it presents local maxima in the directions along the metallic rods (i.e., $\phi = 0°$ and 180°) and minima in the orthogonal ones (i.e., $\phi = 90°$ and 270°). Such response is associated to the field distribution of the nonconservative forces that conform the trap (as the one shown in Fig. 3a-b) and consistent with the local potential barriers found along the strips in Fig. 7. Thus, it is probable that energetic particles scape from these optical traps in the direction across the strips. As wavelength increases further, the metasurface changes its polarization profile and tends to canalize waves along the $\hat{x}$ axis. This mechanism swaps the direction of maximum (minimum) stiffness, which now appears across (along) the strips. In those optical traps, energetic particles will scape in the direction parallel to the strips. For comparison, the trap stiffness obtained focusing the beam over bulk silver and in free-space is shown in Figs. 8b-c. As expected, optical traps engineered over silver only show moderate stiffness around the metal plasmon resonance and always exhibit a symmetrical profile around the trap. Overall, anisotropic metasurfaces significantly boost the stiffness of engineered optical traps over a large frequency band.

## 4. Conclusions

We have put forward the concept of anisotropic and hyperbolic optical traps for the trapping and manipulation of nanoparticles. These optical traps are created by illuminating an anisotropic metasurface with a linearly polarized Gaussian beam and their properties strongly depend on the surface topology and light-matter interactions. To analyse this platform, we have developed a rigorous theoretical formalism able to compute the induced trapping forces based on the anisotropic scattered dyadic Green's function approach merged with the Lorentz force within the Rayleigh approximation. This approach, validated with full-wave numerical simulations in COMSOL Multiphysics, reveals that giant, nonconservative recoil forces pointing towards the beam axis dominate the trap response. These forces appear due to the excitation of ultra-confined surface plasmons on the anisotropic metasurface. Then, we have applied the Helmholtz-Hodge decomposition method to calculate the potential energy of the resulting nonconservative force-field. Our formalism permits to compute, for the first time to our knowledge, fundamental metrics that characterize optical traps engineered over plasmonic materials through nonconservative fields, including spatial potential distribution, trap depth and stiffness, local potential barriers, as well as the minimum laser intensity that achieve stable optical trapping.

The performance of the proposed anisotropic optical traps is outstanding: they exhibit large trap depths over an extremely broadband frequency range that cover the entire visible spectrum and extend well into the infrared band. As a result, a wide variety of low-intensity laser sources can be employed to achieve stable trapping of nanoparticles, avoiding precise alignments between the surface response and the operation wavelength and significantly reducing the possibility of damaging trapped particles due to photoheating. As a specific example, we studied the performance of optical traps engineered over a nanostructured silver layer and analysed how the trap response evolves as the metasurface topology changes from anisotropic elliptical to hyperbolic going through the topological transition. In addition, we found that the momentum imbalance of the SPPs excited by the particle on anisotropic surfaces gives rise to local potential barriers and larger trap stiffness along certain spatial directions, thus breaking the spatial symmetry that characterizes common optical traps. The engineered traps exhibit a much larger potential depth and stiffness than the one found focusing identical Gaussian beam over bulk silver or in free-space, and, more importantly, maintain such response over a large bandwidth. We note that our formalism is based on semi-classical Maxwellian approach and omit additional forces that might originate from other mechanisms, such as Casimir forces [21,63]. Investigating the influence of such forces in the proposed platform is the scope of future research.

Moving forward, ultrathin metasurfaces enable unique possibilities to construct optical traps with excellent performance, including the possibility to engineered local potential barriers, at a desired wavelength, by tailoring the surface topology, local density of states, and the momentum of the supported plasmons. To this purpose, different plasmonic materials – including metals such as gold or silver and semimetals as graphene and WtTe2 [64]– can be appropriately patterned in subwavelength arrangements. In addition, natural anisotropic and hyperbolic materials [65,66] can also be employed to trapping purposes, including hexagonal boron nitride [67], hybrid composites [68,69], van der Waals crystals [37,70-72] and an increasing family of 2D materials [58-61]. We envision that anisotropic and hyperbolic metasurfaces will lead to the next generation of low-power nano-optical tweezers.

This work is supported by the National Science Foundation with Grant No. ECCS-1808400 and a CAREER Grant No. ECCS-1749177.

**References**

[1] Y. Pang and R. Gordon, Optical trapping of a single protein, Nano Lett. **12**, 402-406 (2011).

[2] A. Ashkin and J. M. Dziedzic, Optical trapping and manipulation of viruses and bacteria, Science **235**, 1517-1520 (1987).

[3] A. H. Yang, S. D. Moore, B. S. Schmidt, M. Klug, M. Lipson, and D. Erickson, Optical manipulation of nanoparticles and biomolecules in sub-wavelength slot waveguides, Nature **457**, 71-75 (2009).

[4] Y. Roichman, B. Sun, A. Stolarski, and D. G. Grier, Influence of nonconservative optical forces on the dynamics of optically trapped colloidal spheres: the fountain of probability, Phys. Rev. Lett. **101**, 128301 (2008).

[5] E. Eriksson, J. Enger, B. Nordlander, N. Erjavec, K. Ramser, M. Goksör, S. Hohmann, T. Nyström, and D. Hanstorp, A microfluidic system in combination with optical tweezers for analyzing rapid and reversible cytological alterations in single cells upon environmental changes, Lab Chip **7**, 71-76 (2007).

[6] K. C. Neuman and S. M. Block, Optical trapping, Rev. Sci. Instrum. **75**, 2787-2809 (2004).

[7] I. A. Favre-Bulle, A. B. Stilgoe, E. K. Scott, and H. Rubinsztein-Dunlop, Optical trapping in vivo: theory, practice, and applications, Nanophotonics **8**, 1023-1040 (2019).

[8] D. Gao, W. Ding, M. Nieto-Vesperinas, X. Ding, M. Rahman, T. Zhang, C. Lim, and C. W. Qiu, Optical manipulation from the microscale to the nanoscale: fundamentals, advances and prospects, Light Sci. Appl. **6**, e17039-e17039 (2017).

[9] P. C. Chaumet, A. Rahmani, and M. Nieto-Vesperinas, Optical trapping and manipulation of nano-objects with an apertureless probe, Phys. Rev. Lett. **88**, 123601 (2002).

[10] H. Furukawa and I. Yamaguchi, Optical trapping of metallic particles by a fixed Gaussian beam, Opt. Lett. **23**, 216-218 (1998).

[11] A. Ashkin, J. M. Dziedzic, and T. Yamane, Optical trapping and manipulation of single cells using infrared laser beams, Nature **330**, 769 (1987).

[12] A. Ashkin, J. M. Dziedzic, J. E. Bjorkholm, and S. Chu, Observation of a single-beam gradient force optical trap for dielectric particles, Opt. Lett. **11**, 288-290 (1986).

[13] A. Ashkin, Optical trapping and manipulation of neutral particles using lasers, Proc. Natl. Acad. Sci. **94**, 4853-4860 (1997).

[14] P. C. Chaumet and M. Nieto-Vesperinas, Time-averaged total force on a dipolar sphere in an electromagnetic field, Opt. Lett. **25**, 1065-1067 (2000).

[15] G. Volpe, L. Helden, T. Brettschneider, J. Wehr, and C. Bechinger, Influence of noise on force measurements, Phys. Rev. Lett. **104**, 170602 (2010).

[16] M. G. Silveirinha, S. A. H. Gangaraj, G. W. Hanson, and M. Antezza, Fluctuation-induced forces on an atom near a photonic topological material, Phys. Rev. A. **97**, 022509 (2018).

[17] M. L. Juan, M. Righini, and R. Quidant, Plasmon nano-optical tweezers, Nat. Photonics **5**, 349 (2011).

[18] M. Righini, G. Volpe, C. Girard, D. Petrov, and R. Quidant, Surface plasmon optical tweezers: tunable optical manipulation in the femtonewton range, Phys. Rev. Lett. **100**, 186804 (2008).

[19] Y. Tsuboi, T. Shoji, N. Kitamura, M. Takase, K. Murakoshi, Y. Mizumoto, and H. Ishihara, Optical trapping of quantum dots based on gap-mode-excitation of localized surface plasmon, J. Phys. Chem. Lett. **1**, 2327-2333 (2010).

[20] K. Wang, E. Schonbrun, P. Steinvurzel, and K. B. Crozier, Trapping and rotating nanoparticles using a plasmonic nano-tweezer with an integrated heat sink, Nat. Commun. **2**, 469 (2011).

[21] L. Novotny and B. Hecht, *Principle of nano-optics* (Cambridge, UK, Cambridge University Press, 2012).

[22] F. J. Rodríguez-Fortuño, N. Engheta, A. Martínez, and A. V. Zayats, Lateral forces on circularly polarizable particles near a surface, Nat. Commun. **6**, 8799 (2015).

[23] M. I. Petrov, S. V. Sukhov, A. A. Bogdanov, A. S. Shalin, and A. Dogariu, Surface plasmon polariton assisted optical pulling force, Laser Photonics Rev. **10**, 116-122 (2016).

[24] J. J. Kingsley-Smith, M. F. Picardi, L. Wei, A. V. Zayats, and F. J. Rodríguez-Fortuño, Optical forces from near-field directionalities in planar structures, Phys. Rev. B **99**, 235410 (2019).

[25] A. Ivinskaya, N. Kostina, A. Proskurin, M. I. Petrov, A. A. Bogdanov, S. Sukhov, A. V. Krasavin, A. Karabchevsky, A. S. Shalin, and P. Ginzburg, Optomechanical manipulation with hyperbolic metasurfaces. ACS Photonics **5**, 4371-4377 (2018).

[26] S. B. Wang and C. T. Chan, Lateral optical force on chiral particles near a surface, Nat. Commun. **5**, 3307 (2014).

[27] A. Hayat, J. B. Mueller, and F. Capasso, Lateral chirality-sorting optical forces, Proc. Natl. Acad. Sci. **112**, 13190 (2015).

[28] K. Y. Bliokh, F. J. Rodríguez-Fortuño, F. Nori, and A. V. Zayats, Spin–orbit interactions of light, Nat Photonics **9**, 796 (2015).

[29] J. Petersen, J. Volz, and A. Rauschenbeutel, Chiral nanophotonic waveguide interface based on spin-orbit interaction of light, Science **346**, 67 (2014)

[30] D. O'connor, P. Ginzburg, F. J. Rodríguez-Fortuño, G. A. Wurtz, and A. V. Zayats, Spin - orbit coupling in surface plasmon scattering by nanostructures, Nat. Commun. **5**, 5327 (2014)

[31] N. K. Paul, D. Correas-Serrano, and J. S. Gomez-Diaz, Giant lateral optical forces on Rayleigh particles near hyperbolic and extremely anisotropic metasurfaces, Phys. Rev. B. **99**, 121408 (2019).

[32] J. S. Gomez-Diaz and A. Alù, Flatland optics with hyperbolic metasurfaces, ACS Photonics **3**, 2211-2224 (2016).

[33] J. S. T. Smalley, F. Vallini, S. A. Montoya, L. Ferrari, S. Shahin, C. T. Riley, B. Kanté, E. E. Fullerton, Z. Liu, and Y. Fainman, Luminescent hyperbolic metasurfaces, Nat. Commun. **8**, 13793 (2017).

[34] J. S. Gomez-Diaz, M. Tymchenko, and A. Alù, Hyperbolic metasurfaces: surface plasmons, light-matter interactions, and physical implementation using graphene strips, Opt. Mater. Express **5**, 2313-2329 (2015).

[35] J. S. Gomez-Diaz, M. Tymchenko, and A. Alù, Hyperbolic plasmons and topological transitions over uniaxial metasurfaces, Phys. Rev. Lett. **114**, 233901 (2015).

[36] D. Correas-Serrano, J. S. Gomez-Diaz, M. Tymchenko, and A. Alù, Nonlocal response of hyperbolic metasurfaces, Opt. Express **23**, 29434-29448 (2015).

[37] P. Li, I. Dolado, F. J. Alfaro-Mozaz, F. Casanova, L. E. Hueso, S. Liu, J. H. Edgar, A. Y. Nikitin, S. Vélez, and R. Hillenbrand, Infrared hyperbolic metasurface based on nanostructured van der Waals materials, Science **359**, 892-896 (2018).

[38] A. A. High, R. C. Devlin, A. Dibos, M. Polking, D. S. Wild, J. Perczel, N. P. de Leon, M. D. Lukin, and H. Park, Visible-frequency hyperbolic metasurface, Nature **522**, 192 (2015).

[39] Y. Yermakov, D. V. Permyakov, F. V. Porubaev, P. A. Dmitriev, A. K. Samusev, I. V. Iorsh, R. Malureanu, A. V. Lavrinenko, and A. A. Bogdanov, Effective surface conductivity of optical hyperbolic metasurfaces: from far-field characterization to surface wave analysis, Sci. Rep. **8**, 14135 (2018).

[40] A. Ivinskaya, M. I. Petrov, A. A. Bogdanov, I. Shishkin, P. Ginzburg, and A. S. Shalin, Plasmon-assisted optical trapping and anti-trapping, Light Sci. Appl. **6**, e16258 (2017).

[41] H. Bhatia, G. Norgard, V. Pascucci, and P. T. Bremer, The Helmholtz-Hodge decomposition—a survey, IEEE Trans. Vis. Comput. Graph **19**, 1386-1404 (2012).

[42] www.comol.com

[43] See Supplementary Article Materials for a detailed derivation of the optical trapping forces induced on dipolar Rayleigh particles above anisotropic metasurfaces. Additional details regarding the effective medium theory of anisotropic metasurfaces, particle electromagnetic response and polarization, vertical forces, as well as the influence of the substrate, are also presented.

[44] D. V. Thourhout and J. Roles, Optomechanical device actuation through the optical gradient force, Nat. Photonics **4**, 211 (2010).

[45] J. Kumar, L. Li, X. L. Jiang, D. Y. Kim, T. S. Lee, and S. Tripathy, Gradient force: The mechanism for surface relief grating formation in azobenzene functionalized polymers, Appl. Phys. Lett. **72**, 2096-8 (1998).

[46] A. Rohrbach, Stiffness of optical traps: quantitative agreement between experiment and electromagnetic theory, Phys. Rev. Lett. **95**, 168102 (2005).

[47] L. Novotny, R. X. Bian, and X. S. Xie, Theory of nanometric optical tweezers, Phys. Rev. Lett. **79**, 645 (1997).

[48] M. A. Zaman, P. Padhy, and L. Hesselink, Near-field optical trapping in a non-conservative force field, Sci. Rep. **9**, 1-11 (2019).

[49] S. Sukhov and A. Dogariu, Non-conservative optical forces, Rep. Prog. Phys. **80**, 112001 (2017).

[50] M. A. Zaman, P. Padhy, and L. Hesselink, Solenoidal optical forces from a plasmonic Archimedean spiral, Phys. Rev. A **100**, 013857 (2019).

[51] D. Zill, W. S. Wright, and M. R. Cullen, *Advanced engineering mathematics* (Jones & Bartlett Learning, 2011).

[52] F. Hajizadeh and S. N. S. Reihani, Optimized optical trapping of gold nanoparticles, Opt. Express **18**, 551-559 (2010).

[53] M. Šiler and P. Zemánek, Particle jumps between optical traps in a one-dimensional (1D) optical lattice, New J. Phys. **12**, 083001 (2010).

[54] O. Y. Yermakov, A. I. Ovcharenko, M. Song, A. A. Bogdanov, I. V. Iorsh, and Y. S. Kivshar, Hybrid waves localized at hyperbolic metasurfaces, Phys. Rev. B **91**, 235423 (2015).

[55] I. Trushkov and I. Iorsh, Two-dimensional hyperbolic medium for electrons and photons based on the array of tunnel-coupled graphene nanoribbons, Phys. Rev. B **92**, 045305 (2015).

[56] A. Nemilentsau, T. Low, and G. Hanson, Anisotropic 2D materials for tunable hyperbolic plasmonics, Phys. Rev. Lett. **116**, 066804 (2016).

[57] A. V. Kildishev, A. Boltasseva, and V. M. Shalaev, Planar photonics with metasurfaces, Science **339**, 1232009 (2013).

[58] D. Correas-Serrano, J. S. Gomez-Diaz, A. A. Melcon, and A. Alù, Black phosphorus plasmonics: anisotropic elliptical propagation and nonlocality-induced canalization, J. Opt. **18**, 104006 (2016).

[59] E. Van Veen, A. Nemilentsau, A. Kumar, R. Roldán, M. I. Katsnelson, T. Low, and S. Yuan, Tuning two-dimensional hyperbolic plasmons in black phosphorus, Phys. Rev. Appl. **12**, 014011 (2019).

[60] C. Wang, S. Huang, Q. Xing, Y. Xie, C. Song, F. Wang, and H. Yan, Van der Waals thin films of WTe 2 for natural hyperbolic plasmonic surfaces, Nat. Commun. **11**, 1-9 (2020).

[61] A. J. Frenzel, C. C. Homes, Q. D. Gibson, Y. M. Shao, K. W. Post, A. Charnukha, R. J. Cava, and D. N. Basov, Anisotropic electrodynamics of type-II Weyl semimetal candidate WTe 2, Phys. Rev. B **95**, 245140 (2017).

[62] D. Correas-Serrano, A. Alù, and J. S. Gomez-Diaz, Plasmon canalization and tunneling over anisotropic metasurfaces, Phys. Rev. B **96**, 075436 (2017).

[63] J. L. Garrett, D. A. Somers, and J. N. Munday, Measurement of the casimir force between two spheres, Phys. Rev. Lett. **120**, 040401 (2018).

[64] C. C. Homes, M. N. Ali, and R. J. Cava, Optical properties of the perfectly compensated semimetal WTe 2, Phys. Rev. B **92**, 161109 (2015).

[65] J. Sun, N. M. Litchinitser, and J. Zhou, Indefinite by nature: from ultraviolet to terahertz, ACS Photonics **1**, 293-303 (2014).

[66] S. Guan, S. Y. Huang, Y. Yao, and S. A. Yang, Tunable hyperbolic dispersion and negative refraction in natural electride materials, Phys. Rev. B **95**, 165436 (2017).

[67] J. D. Caldwell, A. V. Kretinin, Y. Chen, V. Giannini, M. M. Fogler, Y. Francescato, C. T. Ellis, J. G. Tischler, C. R. Woods, A. J. Giles, and M. Hong, Sub-diffractional volume-confined polaritons in the natural hyperbolic material hexagonal boron nitride, Nat. Commun. **5**, 1-9 (2014).

[68] S. Dai, Q. Ma, M. K. Liu, T. Andersen, Z. Fei, M. D. Goldflam, M. Wagner, K. Watanabe, T. Taniguchi, M. Thiemens, and F. Keilmann, Graphene on hexagonal boron nitride as a tunable hyperbolic metamaterial, Nat. Nanotech. **10**, 682-686 (2015).

[69] V. W. Brar, M. S. Jang, M. Sherrott, S. Kim, J. J. Lopez, L. B. Kim, M. Choi, and H. Atwater, Hybrid surface-phonon-plasmon polariton modes in graphene/monolayer h-BN heterostructures, Nano Lett. **14**, 3876-3880 (2014).

[70] W. Ma, P. Alonso-González, S. Li, A. Y. Nikitin, J. Yuan, J. Martín-Sánchez, J. Taboada-Gutiérrez, I. Amenabar, P. Li, S. Vélez, and C. Tollan, In-plane anisotropic and ultra-low-loss polaritons in a natural van der Waals crystal, Nature **562**, 557-562 (2018).

[71] Z. Zheng, N. Xu, S. L. Oscurato, M. Tamagnone, F. Sun, Y. Jiang, Y. Ke, J. Chen, W. Huang, W. L. Wilson, and A. Ambrosio, A mid-infrared biaxial hyperbolic van der Waals crystal, Sci. Adv. **5**, eaav8690 (2019).

[72] M. N. Gjerding, R. Petersen, T. G. Pedersen, N. A. Mortensen, and K. S. Thygesen, Layered van der Waals crystals with hyperbolic light dispersion, Nat. Commun. **8**, 1-8 (2017).

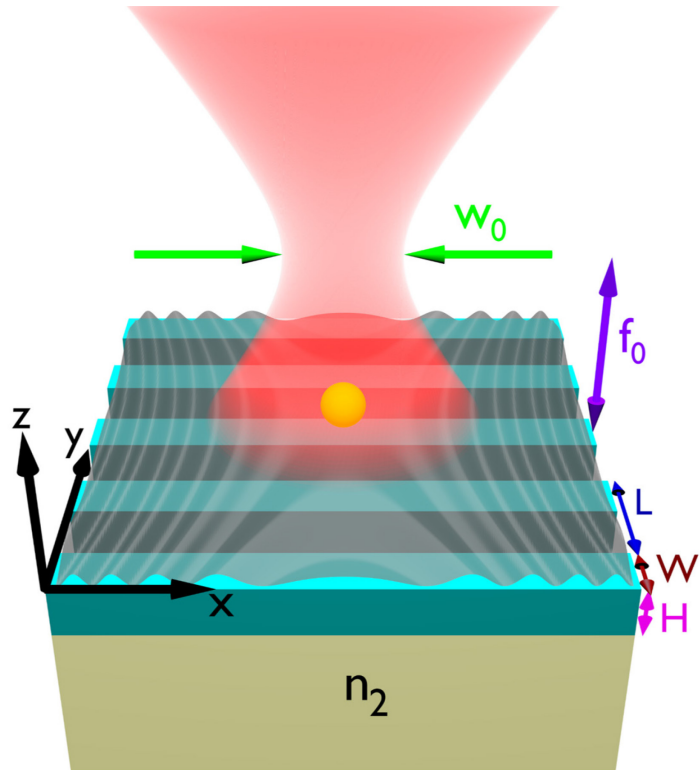

Fig. 1. Hyperbolic optical trap created by illuminating a Rayleigh particle (yellow) located above an ultrathin anisotropic metasurface (cyan) with a p-polarized Gaussian beam (red). The beam has width $w_0$ and has been focused at a distance $f_0$ normal to the surface. During the light scattering process, the particle excites highly confined surface plasmons (grey) on the metasurface propagating away from the beam axis where the optical trap is generated. The hyperbolic metasurface is constructed using subwavelength metallic rods with width W, height H and periodicity L, and is supported by a medium of refractive index $n_2$.

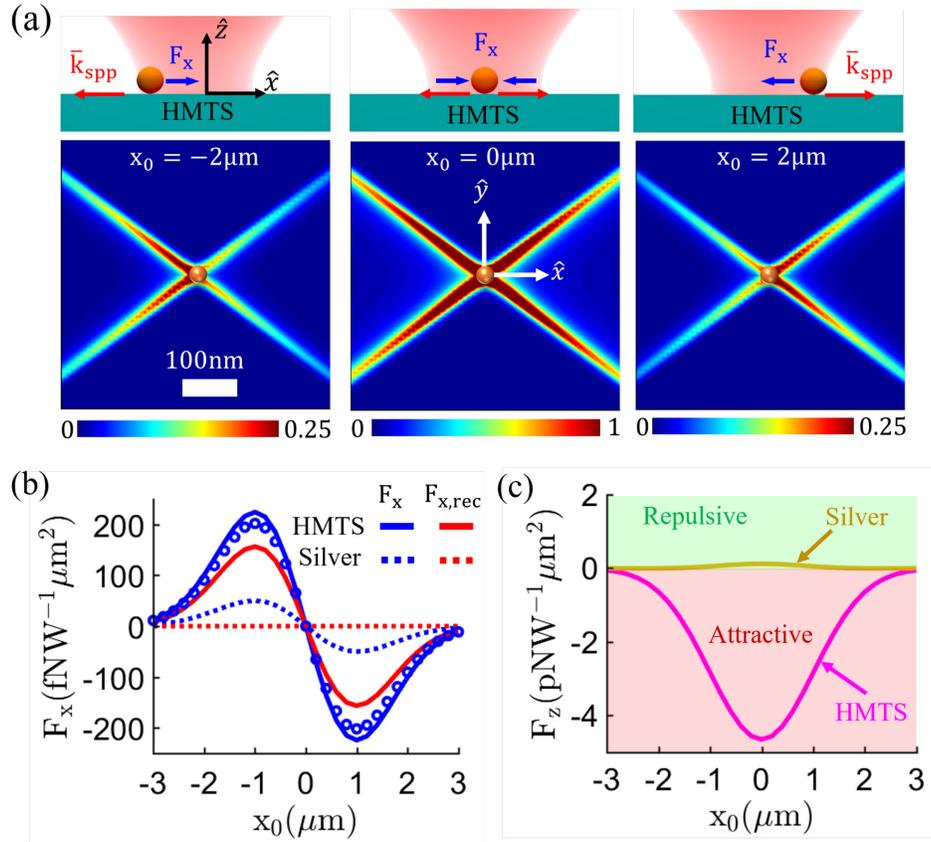

Fig. 2. Trapping Rayleigh particles over a nanostructured metasurface with a Gaussian beam. (a) Normalized power of the surface plasmons excited on the surface when the particle is located in different positions with respect to the beam axis. The top inset illustrates the direction of the plasmon wavevector and the recoil forces acting on the particle. (b) Total lateral forces $F_x$ (blue solid line) and recoil forces $F_{x,rec}$ (red solid line) exerted on the nanoparticle versus its position with respect to the beam axis [43]. Results obtained using COMSOL Multiphysics (markers) are included for validation. Dotted lines correspond to the forces acting on the nanoparticle when the metasurface is replaced with bulk silver. (c) Vertical forces $F_z$ acting on the nanoparticle as a function of its position $x_0$ with respect to the beam axis. Negative (positive) values of $F_z$ correspond to attractive (repulsive) vertical forces toward (away from) the metasurface (bulk silver). The gold nanoparticle has a radius $a = 15$nm and is located in free space at a distance $z_0 = a$ over the metasurface described in Fig. 1 with parameters W = 60nm, L = 180nm, H = 10nm, and $n_2 = 1.05$. The Gaussian beam width is $w_0 = 2\mu m$, focus is $f_0 = 0$, and its operating wavelength is 540nm.

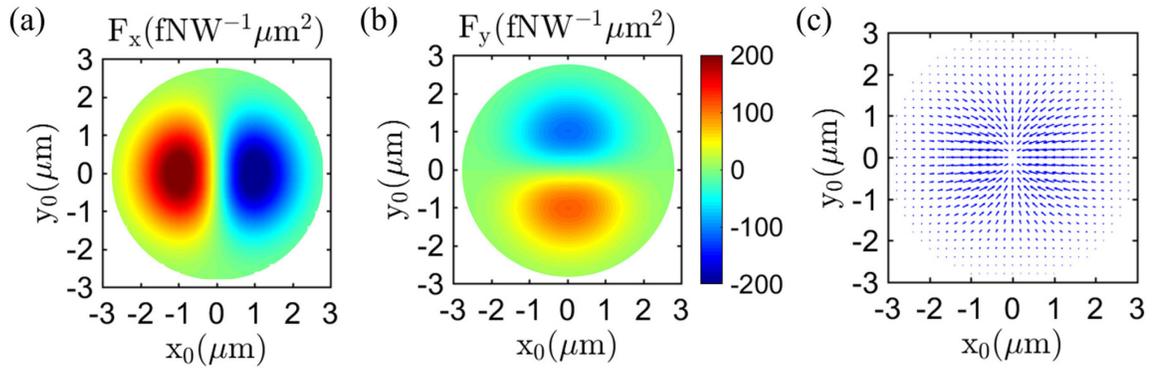

Fig. 3. Optical trapping of a Rayleigh particle located above a hyperbolic metasurface when it is illuminated with a Gaussian beam. (a)-(b) Lateral components of the total force acting on the nanoparticle versus its position $(x_0, y_0)$ with respect to the beam axis. (c) Quiver plot detailing the direction of the lateral forces. Other parameters are as in Fig. 2.

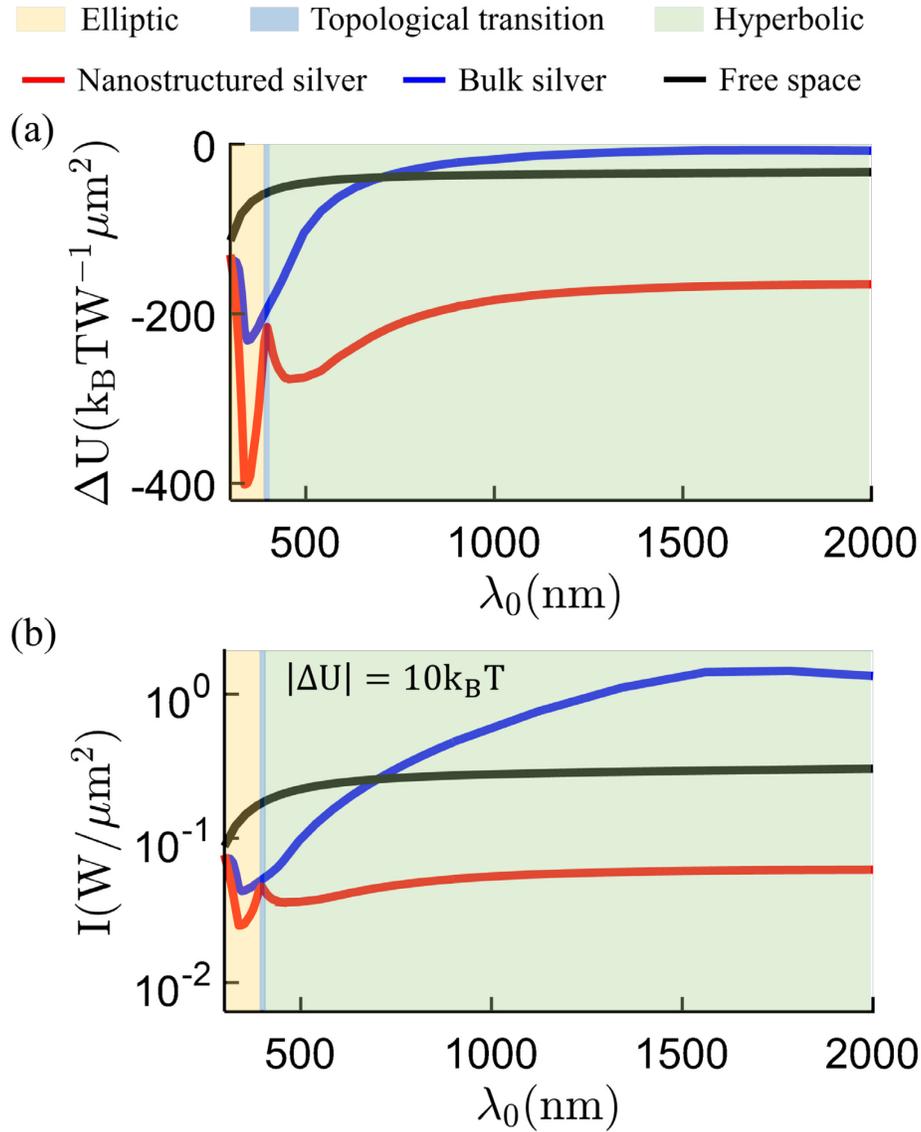

Fig. 4. Performance of optical traps engineered over anisotropic metasurfaces versus frequency. (a) Trap depth normalized with respect to the power density available at the center of the incident Gaussian beam. (b) Minimum amount of power density required to achieve stable trapping. Results are computed for a nanoparticle that is illuminated by a Gaussian beam and is located above an array of silver nanostrips (red), above bulk silver (blue), and in free space (black). The background shaded region corresponds to different metasurface topologies (yellow: elliptic, green: hyperbolic) going through the topological transition (blue) associated with the nanostructured silver. Other parameters are as in Fig. 2.

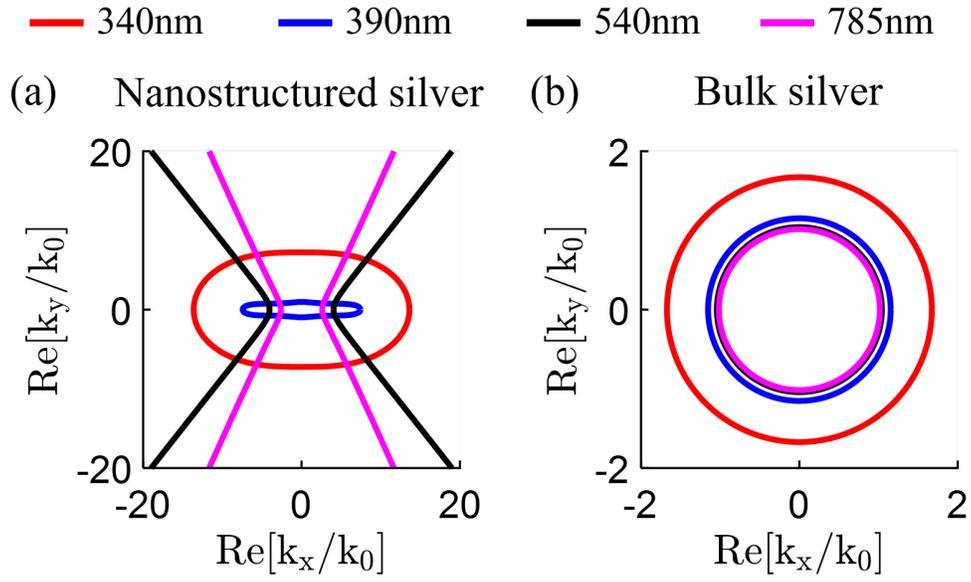

Fig. 5. Isofrequency contour of a nanostructured silver layer (left) and bulk silver (right) at different wavelengths. The physical dimensions of the nanostructure are detailed in Fig. 2.

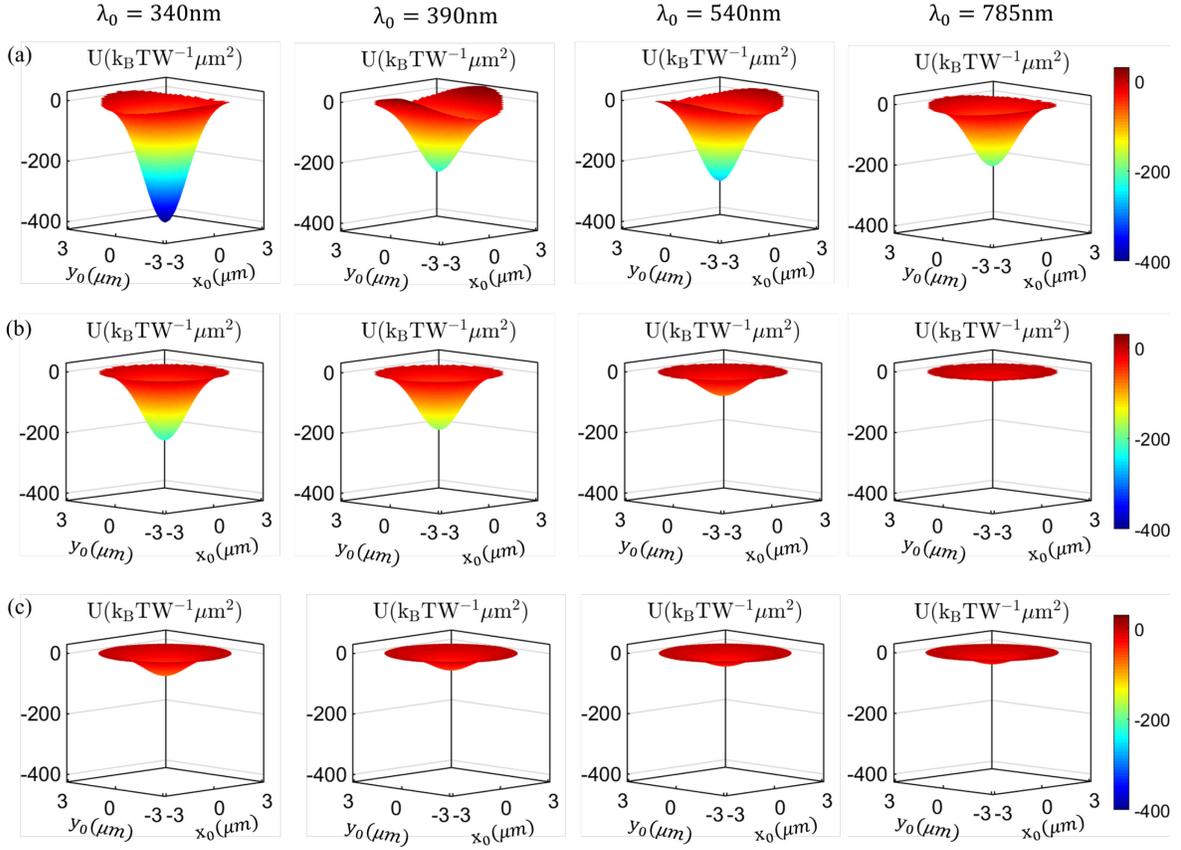

Fig. 6. Trap potential versus the position $(x_0, y_0)$ of the particle when it is illuminated by a Gaussian beam oscillating at 340nm, 390nm, 540nm and 785nm operation wavelength. Results are computed when the particle is located above (a) a nanostructured silver layer, (b) bulk silver, and (c) in free space. Other parameters are as in Fig. 2.

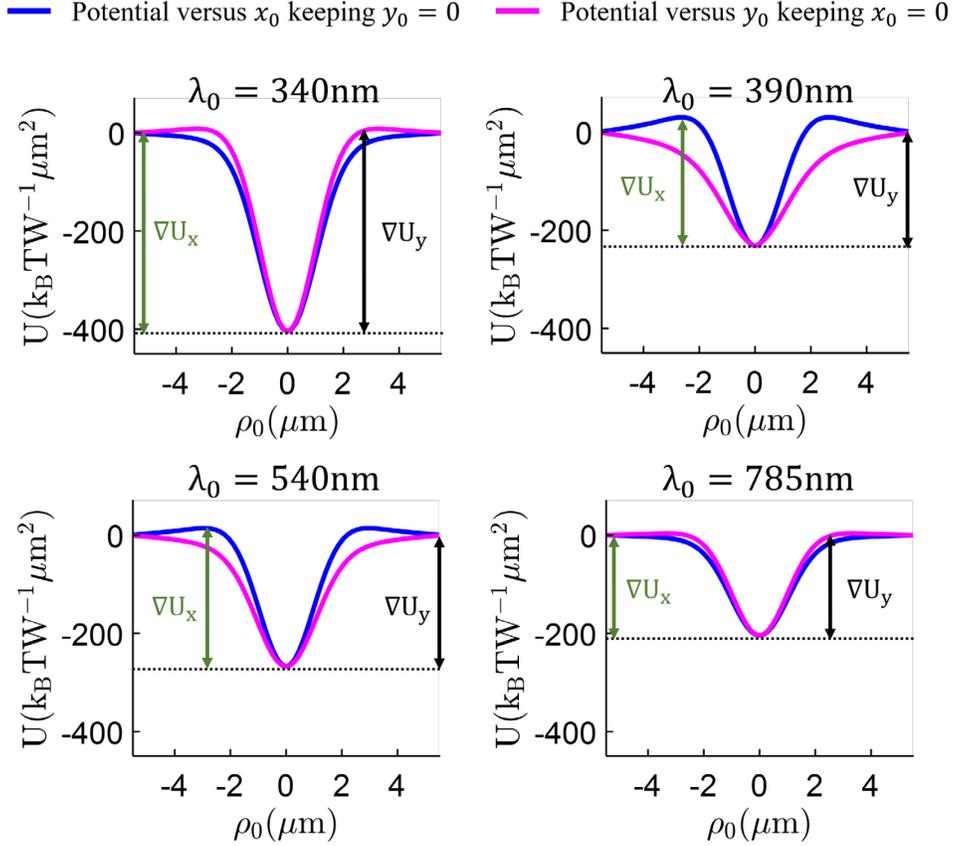

Fig. 7. Trapping potential computed as a function of the particle position $(x_0, y_0)$ along ($x_0$ with $y_0 = 0$; blue line) and across ($y_0$ with $x_0 = 0$; magenta line) the nanorods of a nanostructured silver layer for several operation wavelengths. Local potential barriers along and across the nanorods are denoted as $\Delta U_x$ and $\Delta U_y$, respectively. Other parameters are as in Fig. 2.

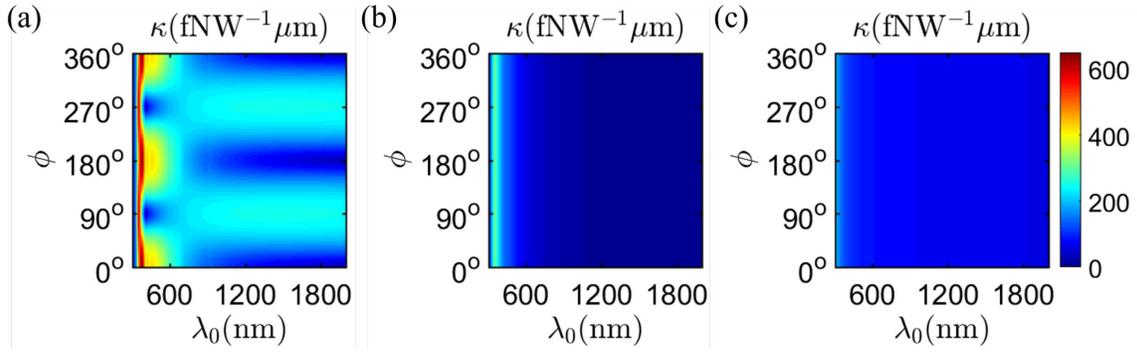

Fig. 8. Trap stiffness induced on a nanoparticle as a function of the wavelength ($\lambda_0$) of the incident Gaussian beam and the polar angle ($\phi$) defined with respect to the $\hat{x}$-axis in Fig. 1. Results are computed for a nanoparticle that is illuminated by a Gaussian beam and is located above a nanostructured silver layer (a), bulk silver (b), and in free space (c). Other parameters are as in Fig. 2.